# The effect of low doses of cadmium and zinc chloride on the Blood-Testis Barrier of Sprague Dawley rats.

Andrea Redondo[1], Raquel Romero[1], Ana Lorenzo[1], Noelia Sanz[2], Esther Durán[1,2], Beatriz Oltra[1,2], José Manuel Pozuelo[1,2], Luis Santamaría[3], and Riánsares Arriazu[1,2*].

[1]Histology Laboratory, Institute of Applied Molecular Medicine, School of Medicine, CEU-San Pablo University, Madrid, Spain.
[2]Department of Basic Medical Sciences, School of Medicine, CEU-San Pablo University, Madrid, Spain.
[3]Department of Morphology, School of Medicine, Autonomous University of Madrid, Madrid, Spain.

**ABSTRACT—Background:** Cadmium chloride is an environmental toxic that affects the male reproductive system. This study was directed 1) to evaluate whether long-term oral exposures to low doses of Cd in rats causes morphological changes, 2) the immunohistochemical TJ and AJ protein expression; and 3) to evidence that zinc exposure can modify Cd effects. **Methods:** Testis of normal rats and rats that have received Cd or Cd+Zn in drinking water during 12 months did not show any morphological change. Immunohistochemistry was used to identify BTB integrity, claudin-1, ZO-1, N-Cadherin and β-catenin were detected by immunohistochemistry and IOD were measure. **Results:** No immunoexpression changes of claudin-1 and ZO-1 were observed. Cd + Zn group showed a significant increase of IOD of N-Cadherin, and a significant decrease of β-catenin and transferrin immunoexpression was observed. **Conclusions:** Ours findings showed that the intake of low oral doses of cadmium chloride over a long time period of time did not cause morphological variation on rat testis. Neither alteration on BTB was found. Zinc chloride was able to strengthen TJ and AJ, and decrease transferrin.

**Key words: Testis prostate, cadmium, zinc, TJ, AJ, immunohistochemistry**

## INTRODUCTION

Cadmium (Cd) is an inorganic toxicant of great environmental and occupational concern which was classified as a human carcinogen in 1993. It is widely dispersed in the environment and causes various diseases, both in occupationally exposed individuals and in the general population. The general population may be exposed to cadmium via contaminants found in food and drinking water, by inhalation of particulates from ambient air or tobacco smoke, or ingestion of contaminated soil or dust. Cd toxicity is associated with damage in different organs, as testes and prostate [1-4]. The rodent testis is extremely sensitive to Cd toxicity. Nevertheless, several factors can modify testicular damage; for example, zinc has been shown that is effective in attenuating the testis damage [5]. A number of authors studied the adverse consequences on testes derived of Cd exposure in many species, using a great range of doses and means of Cd administration [1,6]. The testis is extremely sensitive to Cd toxicity. High levels of Cd causes testicular edema, haemorrhages, necrosis, and sterility in several mammalian species [7,8]. However, long-term toxic effects of low-cadmium exposure in similar conditions to those in the human population are limited.

The blood-testis barrier (BTB) is one of the tightest blood-tissue barriers in the mammalian body. The BTB is a macromolecular tight junction complex generated by somatic Sertoli cells within the seminiferous epithelium. It divides the seminiferous epithelium into the basal and the apical (adluminal) compartments. BTB is not a static ultrastructure, instead; it undergoes extensive restructuring during the seminiferous epithelial cycle of spermatogenesis. The BTB is composed of coexisting tight junctions, ectoplasmic specializations, desmosome-like junctions and gap juctions. All these junctions function together in the maintenance of BTB integrity [9-13].

Protein complexes are very important to ensure the junctions stability. BTB is composed by tight junction (TJ) protein complexes, e.g.claudin/ZO-1, adhesion protein complexes, including adherent junctions (AJ), e.g. N-cadherin/β-catenin, desmosomes, e.g. desmoglein-2/desmocollin-2, and gap junctions, e.g. connexin43/plakophilin-2 [11-13]. The roles of desmosome and gap junction at the BTB remain unknown for decades since their first discoveries in the 1970s [13].

*Correspondence to: R. Arriazu
e-mail: arriazun@ceu.es
Phone: 34-91-3724761; Fax: 34-91-3510496

The purposes of this work were: 1) to evaluate whether long-term oral exposures to low doses of Cd in rats causes morphological changes, 2) the immunohistochemical TJ and AJ protein expression; and 3) to evidence that zinc exposure can modify Cd effects.

## MATERIALS AND METHODS

### Animals and Experimental design

Thirty adult male Sprague-Dawley rats, 30 days old at the beginning of the study, were used. The animals were housed five per cage, with a 12h light-dark cycle. They were supplied with Panlab Lab Chow (Panlab, Barcelona, Spain) and water ad libitum. The experimental protocol followed the guidelines for the care and use of research animals adopted by the Society for the Study of Reproduction.
Sprague Dawley rats were randomly assigned to three groups, according to treatment (10 rats per group). Cadmium chloride (Panreac, Madrid, Spain) was added to the drinking water of the first group at a concentration of 60 ppm, during the time course of the experiment (12 months). The second group received zinc chloride (Panreac, Madrid, Spain) at a concentration of 50 ppm plus cadmium chloride (60ppm) in the drinking water. The third group was used as control and received drinking water that was shown to be free of these metals.

All the animals were euthanized by exsanguination after $CO_2$ narcosis 12 months after the beginning of the experiment. The organic remains from the animals and the residua of drinking water were adequately processed according to the guidelines in relation to the safety in the use of heavy metals established by the communitarian normative of European Union.

### Tissue preparation

All the specimens were fixed by immersion in 4% paraformaldehyde in phosphate buffered saline (PBS) pH 7.4, for 24 hr. After fixation, the testes were cut into 5 mm-wide slices, the section plane perpendicular to the sagittal axis of the gonad. Thereafter, the slices were embedded in paraffin and the paraffin blocks were then serially sectioned at five µm-thicknesses and stained with hematoxylin-eosine or used for immunohistochemical techniques.

### Morphometry

The area of testicular tissue components was determined measuring the area occupied by seminiferous tubules and interstitial, in fifteen fields per animal using MetaMorph software (Leica MMAF 1.4) associated to a Leica Microscope (see in Image analysis) at 400x magnification. The tubular diameter of seminiferous tubules was measured at 100x magnification. For that, 30 tubular rounds profiles or nearly round were ramdomly selected for each animal, and analyzed with MetaMorph software (Leica MMAF 1.4).

### Immunohistochemical Methods

In all the groups, at least five selected slides per animal and per antigen were immunostained. Deparaffinized and rehydrated tissue sections were treated for 30 min with hydrogen peroxide 0.3% in phosphate-buffered saline (PBS) pH 7.4, to block endogenous peroxidase, antigen unmasking was performed with pepsin (15 min, Sigma R2283). To minimize nonspecific binding, sections were incubated with serum blocking solution (Histostain® Bulk Kit, Invitrogen Corporation, Carlsbad, California) and subsequently incubated overnight with primary antibodies in a moist chamber at 4ºC. ZO-1 Rabbit Polyclonal Antibody (1:125, Invitrogen Corporation, Carlsbad, California), Claudin 1 Polyclonal Antibody (1:100, ThermoFisher Scientific Inc.), N-cadherin (D-4) (1:25, SantaCruz Biotechnology Inc.), anti-beta Catenin antibody [E247] (1:250, Abcam, Cambridge, MA), and anti-transferrin antibody (obtained from rabbit, RARa/TRf, Nordic Immunological Laboratories, Teknovas, The Netherland) were used.

The second day, immunohistochemistry was performed with standard procedures using Histostain® Bulk Kit (Invitrogen Corporation, Carlsbad, California). After immunoreactions, sections were counter-stained with Harris hematoxylin, dehydrated in ethanol, and mounted in a synthetic resin (Depex, Serva, Heidelberg, Germany). The specificity of the immunohistochemical procedures was checked by incubation of sections with nonimmune serum instead of the primary antibody.

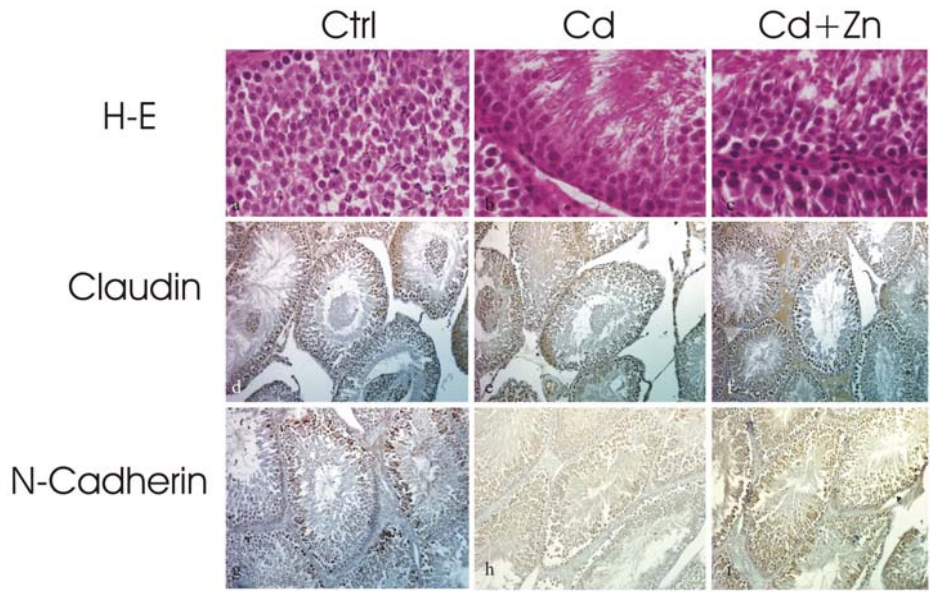

**Figure 1.** Testis from controls and rat exposed to metals. **a-c:** Hematoxylin-Eosin staining (400x). Seminiferous epithelium from a control rat; from a rat exposed to cadmium for 12 months; and from a rat exposed to cadmium plus zinc for 12 months; no histological changes were detected. **d-f:** Claudin-1 immunoexpression (200x). No qualitative changes between groups can be detected. **g-i:** N-Cadherin immunoreactive was detected in all groups studied,

### Image analysis

Digital images of light microscopic were acquired with a Leica DM6000 microscope, Leica Digital Camera DFC425. 10x, 20x and 40x images were captured and processed in TIFF format.

MetaMorph software (Leica MMAF 1.4) was used to determinate de quantitative parameters (morphometric and immunopositivity expression). Immunostaining sections were assessed by estimating the area of the objects and the medium pixel intensity per object as the integrated optical density (IOD). For a better representativeness in the expression levels assessment, a number of 10 images were acquired for each slide and then subjected to densitometric measurements.

### Statistical Analisis

Analysis of the data was done by Student *t*-test for parametric data and Mann Whitney U test for non parametric data. Differences were considered significant when $p < 0.05$ or $p < 0.01$. The software used was GraphPad Prism 5.

### RESULTS

#### Qualitative and morphometry results

No histologic changes were detected in either the Cd or Cd+Zn groups, in comparison with the Ctrl group (figs. 1a-c). Claudin-1 was present in germ cell and in Sertoli cells (fig. 1d-f). ZO-1 was intensely stained in seminiferous tubules in Cd + Zn group. β-catenin is expressed at the apical ectoplasmic specializations and basal compartment in testis seminiferous tubules. The presence of N-cadherin was observed on the surface of spermatogonia and primary spermatocytes (figs. 1g-i). Transferrin immunostaining was observed at the level of Sertoli cells and in germ cells.

The morphometric data are presented in Table 1. No significant difference was found in any of the parameters studied.

#### Immunostaining quantification (IOD)

The IOD of different immunostaining studied are showed in figures 2-4. Added zinc chloride to the drinking water significantly decreased of IOD for β-Catenin in the Cd + Zn group, while Cd group showed values similar to Ctrl group (fig. 2a).

Significantly enhanced of IOD for N-cadherin between Cd + Zn and the other two groups was observed (fig. 2b). Not significant differences were found in IOD between ZO-1 and Claudin-1 of different groups studied (figs. 3a,b).

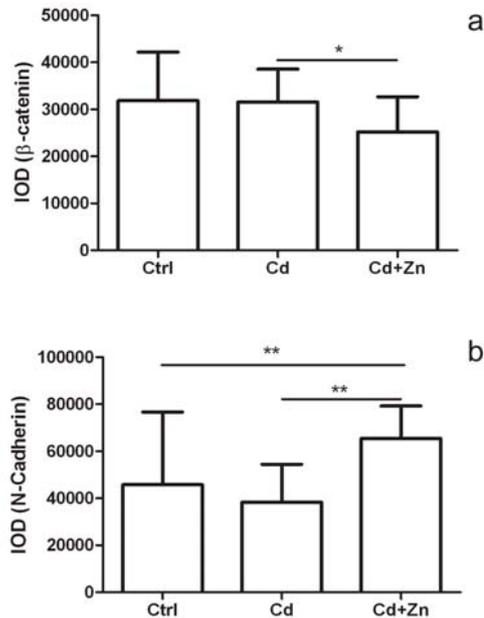

**Figure 2.** Bar graph of the values of the integrated optical density (IOD) of adherent junctions (AJ) in control (Ctrl), cadmium exposed (Cd), and cadmium plus zinc (Cd+Zn)-treated rats. IOD of β-Catenin (a) and N-Cadherin (b) Results are expressed as mean ± SD. * $p<0.05$, ** $p<0.01$

A significant reduction for transferrin occurs in Cd + Zn animals when compared to the Cd and Ctrl groups (fig. 4).

**DISCUSSION**

The protocol of Cd and Cd+Zn exposure employed in this study did not produce histological or quantitative changes in rat seminiferous tubules between treated animals versus control animals. These results are according to the findings of Herranz et al. [14] that reported no significant changes in the tubular volumen or in relation to the amount of germ cells after 12 of cadmium treatment.

Conventionally, apical-basal polarity of epithelial cells is conferred by the differential distribution of tight junction (TJ), between adjacent epithelial cells; behind these are the adherent junctions (AJ) that form the adhesion belt, followed by desmosomes. Collectively, these structures form the junctional complex. Behind the junctional

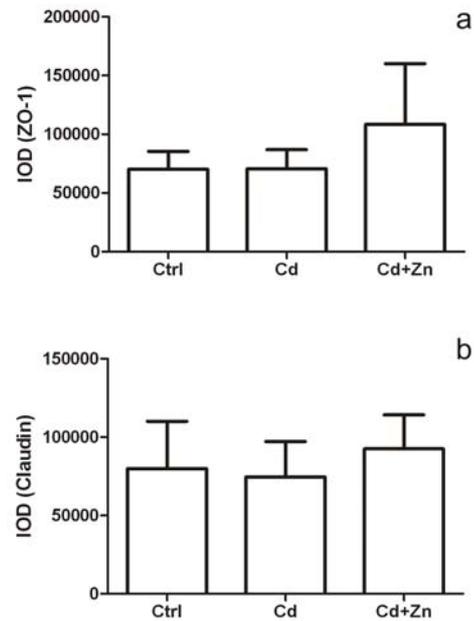

**Figure 3.** Bar graph of the values of the integrated optical density (IOD) of tight junctions (TJ) in different groups studied (Ctrl, Cd, and Cd+Zn-treated rats). IOD of ZO-1 (**a**) and Claudin-1 (**b**) Results are expressed as mean ± SD.

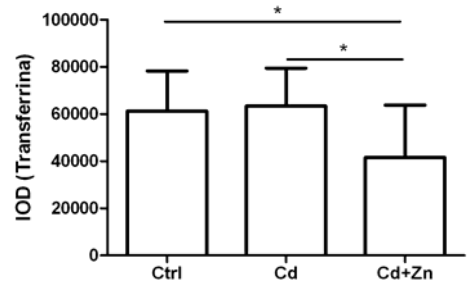

**Figure 4.** Bar graph of the integrated optical density (IOD) of transferrin (Ctrl, Cd, and Cd+Zn-treated rats). Results are expressed as mean ± SD. * $p<0.05$

complex lie the gap junctions [15,16]. The different junction types found in the seminiferous epithelium between two adjacent Sertoli cells and between Sertoli cells and the different classes of developing germ cells are quite different. Sertoli cell TJs are closest to the basement membrane, whereas AJs and desmosome-like junctions are present side by side with TJs and this junctions constitute the BBT, and fuction collectively in the maintenance of BTB integrity [9-13,15,16] which is essential for spermatogenesis and fertility. The tight junction is comprised of several unique proteins, including members of the claudin family, junctional adhesion molecule, the transmembrane protein occludin, linker proteins such as ZO-1, and so on [11,15,17,18]. Claudin-1 is one of the

Table 1. Morphometric parameters of adult rats treated with low doses of cadmium chloride and zinc plus cadmium chloride (mean ± SD).

| Parameter | Ctrl | Cd | Cd + Zn |
|---|---|---|---|
| Tubular diameter (µm) | 533,9 ± 44,92 | 515,5 ± 42,22 | 498,8 ± 56,14 |
| Seminiferous tubule (%) | 81,73 ± 2,967 | 82,24 ± 4,349 | 83,77 ± 2,396 |
| Interstitial tissue (%) | 18,27 ± 2,967 | 17,76 ± 4,349 | 16,23 ± 2,396 |

Ctrl: Control group; Cd: cadmium chloride group; Cd + Zn: cadmium plus zinc chloride group.

most important and critical components in the structural and functional organisation of the tight junctions [19]. ZO-1 is an important cytoplasmic scaffolding protein [20].

To assess the state of the tight junctions at the BTB we determined the changes in the expression of claudin-1 and the tight junction adaptor protein ZO-1, as well as N-cadherin and β-catenin proteins. We did not observe any significant changes between groups in immunohistochemistry quantification of claudin-1 and ZO-1, but Cd+Zn group showed a slightly increase. However, immunoreactivity to N-cadherin and β-catenin showed significant changes in relation to the different groups of treatment. Cd+Zn group was the highest IOD of N-cadherin and the lowest IOD of β-catenin. De Souza et al. (2012) and Elkin et al. (2010) [2,21] detected weight and morphology affectation and BTB disruption only when they administrated high dose of Cd, whereas the histological appearance of the testis remained unaltered as our results showed.

Testicular transferrin, a marker of the integrity of BTB [22], performs essential roles in iron transport and regulation [23]. Mammalian transferrin can also bind various trace metal, including zinc, manganese, and cooper, but the affinity of these trace metals is lower than that for iron [23]. According to some authors [1,24,25], zinc treatment should reduce or abolish the adverse effects of cadmium, given the similarity in the transport characteristics between both metals. In this work we could not observe any transferrin expression difference between Ctrl and Cd groups, but the transferrin immunoexpression was significant reduced in Cd + Zn group. It would be possible that this decrease may be due to increased levels of zinc, as occurs with iron overload [26].

In conclusion, ours findings showed that the intake of low oral doses of cadmium chloride over a long time period of time did not cause morphological variation on rat testis. Neither alteration on BTB was found. Zinc chloride was able to strengthen TJ and AJ, and decrease transferrin.

## AUTHOR CONTRIBUTIONS

Conceived and designed the experiments: RA. Performed the experiments: AR RR AL Analyzed the data: ED BO RA. Contributed reagents/materials/analysis tools: JMP LS RA. Wrote the paper: RA.